\documentclass[aps,pre,floatfix,showpacs]{revtex4-2}
\usepackage{times}
\usepackage{amsmath}
\usepackage{amsfonts}
\usepackage{amssymb}
\usepackage{latexsym}
\usepackage{graphicx}
\usepackage{bm}
\usepackage{verbatim}

\begin{document}

\title{
The Kirkwood-Riseman Model of Polymer Solution Dynamics is Qualitatively Correct
}

\author{George D. J. Phillies}
\email{phillies@wpi.edu, 508-754-1859}

\affiliation{Department of Physics, Worcester Polytechnic
Institute,Worcester, MA 01609}


\begin{abstract}

We use Brownian dynamics to show: For an isolated polymer coil in a shear field, the Kirkwood-Riseman model for chain motion is qualitatively correct.  Under the same circumstances the Rouse model for chain motion is qualitatively incorrect.  The models are qualitatively different.  Kirkwood and Riseman say polymer coils in a shear field perform whole-body rotation; in the Rouse model rotation does not occur. Our simulations demonstrate that \emph{in shear flow}: Polymer coils rotate. Rouse modes are cross-correlated. The amplitudes and relaxation rates of Rouse modes depend on the shear rate. Rouse's calculation only refers to a polymer coil in a quiescent fluid, where there is no viscous dissipation.  Application of the Rouse model to a polymer coil undergoing shear is invalid.

\end{abstract}

\maketitle

\section{Introduction}

Seven decades ago, Kirkwood and Riseman\cite{kirkwood1948a}, Rouse\cite{rouse1953a}, and Zimm\cite{zimm1956a} advanced simple, seemingly transparent models for the dynamics of dilute polymers in solution. A particular focus of their models was a calculation of the polymeric contribution to a solution's viscosity.  The models were similar in that each approximated a polymer coil as a series of hydrodynamically active points ("beads") held together by hydrodynamically inert connectors ("springs").

Here the similarity between the models ends. As is not uniformly recognized, the Kirkwood-Riseman and Rouse-Zimm models give entirely contradictory descriptions of how polymer chains move in solution. In the Rouse and Zimm models, the connectors ("springs") between the beads create Hooke's-law restoring forces that pull the beads together.  Thermal fluctuations in the solvent create random forces on the polymer beads, tending on the average to drive the beads apart. The competition between the spring and thermal forces determines the size of a polymer coil.  Rouse-Zimm polymer coils can translate; all other bead motions are described by internal modes in which the relative positions of the beads change. These internal modes are claimed to be responsible for the polymer coil's contribution to the viscosity.

In contrast, in the Kirkwood-Riseman model, the dynamics of a polymer coil are approximated as comprising uniform translation and whole-body rotation.  The distances between polymer beads are approximated as having their average values.  According to Kirkwood and Riseman, the dominant contributions to a polymer's intrinsic viscosity and diffusion coefficient arise from whole-body translation and rotation. While Kirkwood and Riseman recognized that polymer beads can move with respect to each other, so that polymer chains do have segmental dynamics, within their model these internal motions were neglected.

Polymer models make different assumptions as to the role of hydrodynamic interactions between different parts of the polymer chain. In free-draining models, such as the Rouse model, bead-bead hydrodynamic interactions are absent.  In non-draining models, such as the Zimm model, polymer beads have hydrodynamic interactions, described in the Zimm model by the Oseen hydrodynamic interaction tensor. The Kirkwood-Riseman model has two parts: In the calculation of the intrinsic viscosity, the chain is taken to be non-draining.  In the calculation of how the chain moves in response to an applied shear field, the model may be equally said to be free-draining or non-draining, because Newton's Third Law of Motion guarantees that internal (here, hydrodynamic) forces  have no effect on the model's motions within the constraints of the model.

This paper presents a simulational test of the Rouse and Kirkwood-Riseman descriptions of polymer motion. We examine an isolated chain in a hydrodynamic shear field.  We employ a wider range of computational diagnostics than has sometimes been used in the past to interpret polymer dynamics.  In particular, we ask whether or not a chain in a shear field rotates.  We ask if fluctuations in Rouse mode amplitudes are cross-correlated, and if Rouse mode amplitudes or relaxation rates depend on the applied shear rate.  We demonstrate alternative diagnostics for studying polymer motion.

The following Section of this paper outlines salient features of the Rouse and Kirkwood-Riseman models.  A further Section describes our simulation procedures, including the physical quantities that we calculated. We then outline our major results, revealing the relative validities of the Rouse and Kirkwood-Riseman models as descriptions of polymer dynamics during a rheological experiment. To anticipate our results, we show that Kirkwood and Riseman are correct, and Rouse was incorrect.

\section{Rouse and Kirkwood-Riseman Models}

This section present aspects of the Rouse\cite{rouse1953a} and Kirkwood-Riseman\cite{kirkwood1948a} models. We begin with the more familiar Rouse model, and then consider the Kirkwood-Riseman model.

Rouse's original treatment was quite involved.  As is often the case with novel theoretical results, as time advances the key aspects of the calculation are abstracted from the original structure. The presentation of Doi and Edwards\cite{doi1986a} and the more extended development by Padding\cite{padding2005a} are followed here. The Rouse model describes an isolated polymer in a solvent. The polymer is approximated as a line of $N$ beads, each linked to the next by a springlike connector. The bead positions are denoted $(\bm{R}_{1}, \bm{R}_{2}, \ldots \bm{R}_{N})$. The beads are points having no excluded volume; they are all free to move with respect to each other. Each bead has a hydrodynamic drag coefficient $f$.  The connectors do not interact with the solvent.

The strength of the connectors is determined by the Gaussian statistics that describe the shape of a random-walk polymer coil. In Rouse's model, each bead represents some substantial number of monomers. The distance along the polymer chain from each bead to the next is sufficiently large that the bead-bead distances $r_{i,i+1} = |\bm{R}_{i+1} - \bm{R}_{i}|$ have Gaussian distributions $P(r_{i,i+1}) \sim \exp(- \alpha r_{i,i+1}^{2})$.

Rouse implicitly explains that for each statistico-mechanical distribution function $P(r_{ij})$, there is a corresponding potential of average force $W(r_{ij})$, namely
\begin{equation}
       W(r_{ij}) = -k_{B} T \ln(P(r_{ij})).
       \label{eq:PoAF}
\end{equation}
Here $k_{B}$ is Boltzmann's constant and T is the absolute temperature. The potential of average force gives the average force between two adjoining beads that are a distance $r_{ij}$ apart.

The calculations here use Rouse's original potential of average force
\begin{equation}
\label{eq:rousepotential}
     W(r_{ij}) = \frac{1}{2} k r_{ij}^{2},
\end{equation}
for two beads $i$ and $j$  that adjoin along the polymer chain. In Rouse's model, the force constant $k$ is determined by the mean-square bead separation $b$, namely
\begin{equation}
\label{eq:kdefinition}
  k = \frac{3 k_{B} T}{b^{2}}.
\end{equation}
It is possible to use considerably more sophisticated forms for the potential energy of the polymer chain.  Note, for example, work of Tsalikis, et al.\cite{tsalikis2017a}, Perez-Aparicio, et al.\cite{perez2011a}, and Kalathi, et al.\cite{kalathi2014a}. The objective here is to test the Rouse and Kirkwood-Riseman models. To test the Rouse model, we must use Rouse's potential. The Kirkwood-Riseman model specifies only average interbead distances, and does not invoke a particular form for the interbead potential energy.

In the original Rouse model, a bead $i$ was also subject to a thermal force $\mathbf{\cal{F}}_{i}(t)$ due to fluctuations in the solvent. Hydrodynamic interactions between beads, and correlations between the thermal forces on different beads, were neglected in the Rouse model but included in the elsewise-similar Zimm model.

We can now write the equations of motion -- Newton's second law -- for each bead of the Rouse model. The drag force on each bead is large. On the time scales of interest bead motions are massively overdamped. Bead inertia is therefore neglected. If bead inertia vanishes, the total force on each bead must also vanish. The direct forces (the spring forces) on each bead must therefore cancel the hydrodynamic forces. The equations of motion for beads other than the two end beads (beads 1 and $N$) are then
\begin{equation}
    \label{eq:rousebeads1}
     f \frac{d \bm{ R}_{i}}{dt } = -k(2\bm{ R}_{i} - \bm{ R}_{i-1} - \bm{ R}_{i+1})  + \mathbf{\cal{F}}_{i}(t),
\end{equation}
while for the first and last beads in the chain one has
\begin{equation}
    \label{eq:rousebeads2}
     f \frac{d \bm{ R}_{1}}{dt } = -k(\bm{ R}_{1} - \bm{ R}_{2})  +  \mathbf{\cal{F}}_{1}(t),
\end{equation}
and
\begin{equation}
    \label{eq:rousebeads2N}
     f \frac{d \bm{ R}_{N}}{dt } = -k(\bm{ R}_{N} - \bm{ R}_{N-1})  +  \mathbf{\cal{F}}_{N}(t).
\end{equation}
It is generally the case that the bead positions $\bm{R}_{i}$ are not all equal to each other.  As a result, the spring forces on individual beads are not zero, so in general the beads must be moving with respect to the solvent to create the countervailing hydrodynamic forces.

The above are $N$ vector equations. They correspond to a total of $3N$ scalar equations describing bead motions.  The direction cosine for the $x$-component of the force between beads $i$ and $i+1$ is $(x_{i+1} - x_{i}) /|\bm{ R}_{i+1} - \bm{ R}_{i}|$.  Corresponding forms give the $y$ and $z$ direction cosines for each force vector.  The $N$ vector equations given above therefore correspond to $3N$ scalar equations such as
\begin{equation}
    \label{eq:rousebeads3}
     f \frac{d x_{i}}{dt } = -k(2x_{i} - x_{i-1} - x_{i+1})  + {\cal F}_{xi},
\end{equation}
$x_{i}$ being the $x$-component of the bead coordinate of bead $i$ and ${\cal F}_{xi}$ being the $x$ component of the thermal force on bead $i$.   As explained by Rouse in his original paper, the equations for the $x$ coordinates, for the $y$ coordinates, and for the $z$ coordinates are, except for the coordinate label, the same as each other. Changing the $x$-component of a particle's position has no effect on the $y$ and $z$ components of the forces on any particle, and correspondingly for displacements of a bead in the $y$ or $z$ directions. The equations of motion for the $x$, $y$, and $z$ coordinates are thus completely uncoupled. The equations of motion therefore partition into three sets of $N$ coupled equations, one set for each of the three coordinate axes. Because each set of equations is the same as the others, except for the label on the coordinates, only one set of $N$ equations needs to be solved. The solutions for the other two sets of equations can be obtained by a change of the coordinate label.  While the equations of motion of the beads do partition into three sets of $N$ equations, one set for each dimension, the model is three-dimensional, not one-dimensional.  Beads move in all three coordinate directions.

Equation \ref{eq:rousebeads3} and the matching equations for beads $1$ and $N$ are a set of $N$ coupled linear differential equations whose coefficients are constants. The solutions are therefore a set of $N$ eigenmodes $Q_{n}$ describing motions parallel to one of the three coordinate axes, each mode having a corresponding eigenvalue $\Gamma_{n}$.  One mode has eigenvalue $\Gamma_{0} =0$; that mode corresponds to uniform translation of all beads in the same direction. The other $N-1$ modes decay exponentially ($\exp(- \Gamma_{n} t)$) in time; their relaxation rates $\Gamma_{n}$ are
\begin{equation}
      \label{eq:relaxationtimes}
   \Gamma_{n} = \frac{8 k \sin^{2} (n \pi/2N)}{f}
\end{equation}
with $n \in (1, N-1)$ being the mode label.

The normal mode amplitudes $C_{xn}(t)$ for the $x$-coordinate modes may be calculated from the bead coordinates $x_{i}(t)$ via
\begin{equation}
    \label{eq:modeamplitudes}
     C_{xn}(t)  = \frac{1}{N} \sum_{i=1}^{N} x_{i}(t) \cos\left(\frac{n \pi (i-1/2)}{N}\right).
\end{equation}
Entirely similar equations give the amplitudes $C_{yn}$ and $C_{zn}$ of the $y$- and $z$-coordinate modes. The inverse equations give the $x_{i}$ in terms of the normal mode amplitudes as
\begin{equation}
   \label{eq:beadpositions}
   x_{i}(t) = C_{x0}(t) + 2 \sum_{n=1}^{N-1} C_{xn}(t) \cos\left(\frac{n \pi (i-1/2)}{N}   \right).
\end{equation}
Totally similar equations describe the $y$ and $z$ modes.  Standard mathematical techniques show how the random forces ${\cal F}_{xi}(t)$ serve as source terms, driving the fluctuations in the $C_{xn}(t)$.

There are three coordinate axes, so the relaxation rates $\Gamma_{n}$ are three-fold degenerate.  For each $n$, the same relaxation rate applies to all three coordinate axes. The modes having degenerate eigenvalues are orthogonal; their amplitudes fluctuate independently. The Rouse model thus has three translational modes, each with eigenvalue zero, and $3N-3$ internal modes ('internal' in the sense that in each internal mode the beads move with respect to each other as time goes on) having non-zero eigenvalues.

On setting all but one of the $C_{xi}$ to zero, eq.\ \ref{eq:beadpositions} gives the representation in bead position space of the eigenvector corresponding to $C_{xi}$. The Rouse eigenvectors thus provide a set of $3N$ normal coordinates that can replace the bead coordinates $\{x_{i}, y_{i}, z_{i}\}$ as a specification of the polymer's configuration. Eqs.\ \ref{eq:modeamplitudes} and \ref{eq:beadpositions} may be interpreted as a pair of discrete Fourier transforms, in which $i-1/2$ plays the role of the position coordinate, $n \pi/N$ is the wave vector, and $x_{i}$ and $C_{xn}$ are the amplitudes of the function and its transform at $i$ and $n$, respectively.

Rouse uses the Rouse modes to describe the behavior of a polymer coil in a shear flow.  In Rouse's calculation, a shear flow exerts forces on the polymer.  The polymer's responses are described by the Rouse modes.  The Rouse solutions therefore were taken by Rouse to be valid descriptions of polymer motion when a shear flow is applied.

Polymer coils whose motions are described by Rouse's model have one ill-recognized property: They do not rotate. This property follows by comparison with a standard problem in classical mechanics, namely the vibrational modes of an isolated molecule.  In general, an $N$-atom molecule has 3 translational modes with eigenvalue zero, 3 rotational modes with eigenvalue zero, and $3N-6$ internal vibrational modes. The internal modes are the modes that change the distances between pairs of atoms.  In translation and rotation the distances between the atoms remain fixed. The Rouse problem only differs from the molecular vibration problem in that the Rouse equations of motion are overdamped, so the Rouse amplitudes relax exponentially at some rate $\Gamma_{n}$ rather than oscillating at some frequency $\omega_{n}$.  A polymer coil is therefore like a vibrating isolated molecule in having a total of $3N$ modes.  However, the $3N$ modes of the Rouse model include 3 translational modes and $3N-3$ internal modes, for a total of $3N$ modes, leaving no modes available for rotational motion.

The statement that Rouse chains do not rotate is not new. Rouse specifies in his paper that a polymer coil under shear does not rotate, namely (his paper, p. 1274, column 2) "...since the velocity of the liquid has a nonvanishing component only in the $x$ direction, the components $(\dot{y}_{j})_{\alpha}$ and $(\dot{z}_{j})_{\alpha}$ are zero."  $(\dot{y}_{j})_{\alpha}$ and $(\dot{z}_{j})_{\alpha}$ are the velocities of bead $j$ in the $y$ and $z$ directions due to the shear.  If the chain is rotating, either $(\dot{y}_{j})_{\alpha}$ or $(\dot{z}_{j})_{\alpha}$  must be non-zero.  Rouse also argues his paper (p. 1274, column 2, top) that '...an atom at the junction between two submolecules...' (springs) moves '...with a velocity equal to that of the surrounding liquid...' except for Brownian motion, because, according to Rouse, otherwise there would be motion of the solvent relative to the polymer chain, leading to energy dissipation. If the beads only move with the liquid, then they can only be moving parallel to the $x$-axis.

We now consider the Kirkwood-Riseman model. While both models refer to a line of beads, the Kirkwood-Riseman model is radically different from the Rouse model. The Kirkwood-Riseman  model is based on three fundamental assumptions. First, all distances between pairs of beads are treated as being their statistico-mechanical average values; fluctuations and changes in these distances are explicitly not included in the model.  Second, the distribution of beads around the chain center-of-mess is spherically symmetric. Third, the system is massively overdamped, so that the inertia of the polymer coil is negligible.  These three assumptions completely define the chain dynamics, the description of how a Kirkwood-Riseman polymer chain moves in solution. Kirkwood and Riseman recognized that a polymer coil has internal modes ("fluctuations") so that polymer beads actually do move with respect to each other, but these bead motions were specified as being not included in their model.

The system is heavily overdamped, so its inertia is negligible.  The total force on the chain must therefore be zero.  The moments of inertia of the chain are negligible.  The total torque on the chain must therefore also be zero. The chain satisfies these two zero conditions by adjusting its linear velocity $\bm{V}$ and its angular velocity $\bm{\Omega}$
until the total force and the total torque on the chain both vanish.

Kirkwood and Riseman consider how a polymer coil moves in a shear field in which the fluid velocity is
\begin{equation}
       \label{eq:kirkwoodshear}
      \bm{u}_{i} = \bm{u}_{i}^{(0)} + G y_{i} \bm{\hat{i}},
\end{equation}
Here $\bm{u}_{i}^{(0)}$ is a possible uniform motion of the fluid, $G$ is a constant linear shear gradient, $y_{i}$ is the $y$-component of the vector location of bead $i$, and $\bm{\hat{i}}$ is the unit vector parallel to the $x$-axis.  We do not follow Kirkwood's notation closely.  Kirkwood and Riseman included in their model bead-bead hydrodynamic interactions as described by the Oseen tensor. We return to these internal interactions below.

The Kirkwood-Riseman model describes a chain of $N$ beads having coordinates $(\bm{R}_{1}, \bm{R}_{2}, \ldots \bm{R}_{N})$. Sequential bead positions form a highly restricted random walk.  Each bead is subject to a hydrodynamic force $\bm{F}_{iH}$ exerted by the fluid, and to forces due to the bonds connecting that bead to its neighbors along the chain.  $\bm{F}_{iH}$ is determined by the bead drag coefficient $f$, the velocity $\bm{v}_{i}$ of the bead, and the velocity $\bm{u}_{i}$ that the fluid would have had, at the location of bead $i$, if bead $i$ were not there, via
\begin{equation}
      \label{eq:beaddragforce}
     \bm{F}_{iH} =  f(\bm{u}_{i} - \bm{v}_{i}).
\end{equation}

$\bm{F}_{iH}$ is the hydrodynamic force on the bead, not the total force.  The total force on each bead, including the forces due to links to adjoining beads, vanishes, so $\bm{F}_{iH}$ in general is non-zero. Correspondingly, the bead and solvent velocities are in general not equal to each other.

In the Kirkwood-Riseman model, the velocity of bead $i$ is
\begin{equation}
     \bm{v}_{i} = \bm{V} + \bm{\Omega} \times \bm{s}_{i}.
     \label{eq:beadvelocity}
\end{equation}
Here $\bm{s}_{i}$ is the vector from the chain center-of-mass to bead $i$, $V$ is a linear velocity, the same for each bead, and $\Omega$ is an angular rotation rate, the same for each bead.  Internal modes neglected in the model would add to the right hand side of this equation an additional term $\dot{\xi}_{i}$, the contribution of the internal modes to the bead velocity; that term does not appear in the model.

Kirkwood and Riseman use the zero-total-force and zero-total-torque conditions to determine $\bm{V}$ and $\bm{\Omega}$ in terms of $\bm{u}_{i}^{(0)}$ and $G$, finding
\begin{equation}\label{eq:Vvalue}
  \bm{V} =  \bm{u}_{i}^{(0)} + G Y_{0} \bm{\hat{i}}
\end{equation}
and
\begin{equation}\label{eq:Omegavalue}
   \bm{\Omega} = - \frac{G}{2}\bm{\hat{k}}.
\end{equation}
$Y_{0}$ is the y-coordinate of the polymer chain's center of mass, and $\bm{\hat{k}}$ is the unit vector in the $z$-direction. The model predicts viscous dissipation because the bead velocity $\bm{v}_{i}$ and the solvent velocity $\bm{u}_{i}$ cannot be equal at every point.  For example, for most beads $\bm{v}_{i}$, but not $\bm{u}_{i}$, will have a non-zero $y$-component.

The calculations leading to eqs.\ \ref{eq:Vvalue} and \ref{eq:Omegavalue} make no reference to hydrodynamic interactions between polymer beads.  To calculate viscous dissipation, Kirkwood and Riseman then insert bead-bead hydrodynamic interactions.  However, bead-bead hydrodynamic interactions have no effect on the dynamic model specified by eqs.\ \ref{eq:beadvelocity}-\ref{eq:Omegavalue}. The dynamic model is not affected by intrachain hydrodynamic interactions because bead-bead hydrodynamic interactions are \emph{internal} forces, forces between different beads on the same chain. The total force and the total torque exerted on a polymer chain by bead-bead forces must both vanish, an outcome guaranteed by Newton's Third Law of Motion. Adding hydrodynamic interactions has no effect on the motions described by eqs.\ \ref{eq:beadvelocity}-\ref{eq:Omegavalue}, these being the equations that completely specify the Kirkwood-Riseman dynamic model.

In applying the Oseen tensor to describe bead-bead hydrodynamic interactions, Kirkwood and Riseman took the distance between each pair of beads to be the equilibrium average distance between those two beads.  Fluctuations in those interbead distances, and the time dependences of those fluctuations, were approximated by Kirkwood and Riseman as not being important.

\section{Computational}

A Brownian dynamics simulation for a polymer in a shear field was implemented. The equations of motion resemble the Rouse equations of motion, but a shear field has been added.  The simulations reveal how a Rouse-like polymer coil moves in a rheological experiment, as envisioned by Rouse and also by Kirkwood and Riseman.

For beads $i \in (2,\ldots, N-1)$, we write these as
\begin{equation}
    \label{eq:rousebeads4}
      \frac{d \bm{ R}_{i}}{dt } = f^{-1} (-k(2\bm{ R}_{i} - \bm{ R}_{i-1} - \bm{ R}_{i+1})) + G y_{i} \mathbf{\hat{i}}   + \mathbf{\cal{F}}_{i}(t),
\end{equation}
$\bm{\hat{i}}$ being the unit vector in the $x$-direction. For beads $1$ and $N$ the equations of motion are
\begin{equation}
    \label{eq:rousebeads4a}
      \frac{d \bm{ R}_{1}}{dt } = f^{-1} (-k(\bm{ R}_{1} - \bm{ R}_{2})) + G y_{1} \bm{\hat{i}}   + \mathbf{\cal{F}}_{1}(t),
\end{equation}
and
\begin{equation}
    \label{eq:rousebeads4b}
      \frac{d \bm{ R}_{N}}{dt } = f^{-1} (-k(\bm{N}_{i} - \bm{ R}_{N-1})) + G y_{N} \bm{\hat{i}}   +\mathbf{\cal{F}}_{N}(t).
\end{equation}
These equations describe chain motion relative to the chain center. The shear force is directed in the $\bm{\hat{i}}$ direction, and changes linearly with the distance in the $y$-direction from the chain center-of-mass.  The thermal forces $\mathbf{\cal{F}}_{i}(t)$ were generated using standard methods as independent random variables having Gaussian distributions. The bead displacements during a single time step $\Delta t$ are $\Delta t \frac{d \bm{ R}_{i}}{dt}$. The forces are re-evaluated after each time step to compute the trajectory of each bead.

Multiple characteristic functions of chain behavior were determined. Most of these functions were used as diagnostics to validate the core software. The radius of gyration and mean-square radius of gyration were calculated.  The mean-square center-of-mass displacement was found to be linear in time, as expected.  The second $\langle x_{\alpha} x_{\beta} \rangle$ and fourth  $\langle x_{\alpha}^{2} x_{\beta}^{2} \rangle$ moments of the bead distribution around the center of mass were calculated.  Here $\alpha, \beta \in (1,3)$ designate individual cartesian components of the vectors from the chain center to each bead, the average being over all beads and all times.

Distribution functions for the nearest-neighbor distance, the second-nearest-neighbor distance, the magnitude of the end-to-end vector, the distance from the polymer center-of-mass to each bead, and the distances between all pairs of beads were measured.   The time autocorrelation functions $\langle \bm{R}_e(0) \cdot \bm{R}_e (t) \rangle$  and   $\langle \bm{\hat{R}}_e(0) \cdot \bm{\hat{R}}_e (t) \rangle$   of the chain end-to-end vector $\bm{R}_e = \bm{R}_{N} - \bm{R}_{1}$ and its unit vector $\bm{\hat{R}}_e = (\bm{R}_{N} - \bm{R}_{1})/\mid \bm{R}_{N} - \bm{R}_{1})\mid$ were obtained.

Using eq.\ \ref{eq:modeamplitudes}, we calculated the time-dependent Rouse amplitudes $C_{\alpha, n}(t)$ of the bead positions.  For an $N$-bead system there are $3(N-1)$ such components, plus the three $C_{\alpha,0}$ describing the polymer center of mass position. We also calculated several time-dependent spatial Fourier components
\begin{equation}
    \label{eq:fouriercomponents}
    a_{k,\alpha}(t) = \sum_{i=1}^{N} \cos( k r_{i\alpha}(t))
\end{equation}
of the bead locations.  Here $k$ is the wavenumber for the transformation and $r_{i\alpha}(t)$ is the $\alpha^{\rm th}$ Cartesian component of the location of bead $i$, relative to the chain center of mass, at time $t$.

Finally, we calculated time-dependent Haar-like\cite{haar1910a} wavelet\cite{daubechies1992a} components $c(n,\alpha,j)(t)$ and $d(n,\alpha,j)(t)$ of the particle positions.  In this calculation, $n$ is the wavelet decomposition level, $\alpha$ is again the Cartesian coordinate, and $j$ labels the wavelet location along the polymer chain. The maximum value of $j$ depends on $N$.  For a $2^{m}$ bead polymer the upper limit on $j$ is $2^{m-n}$ with $m-n \geq 0$.  The decomposition proceeds naturally if for some integer $m$ there are $2^{m}$ beads in the chain. The wavelet components are defined
\begin{eqnarray}
 \nonumber 
  c(1,\alpha,j) &=& (r_{\alpha, 2*j} + r_{\alpha, 2*j-1})/2  \\
  d(1,\alpha,j) &=& (r_{\alpha, 2*j} - r_{\alpha, 2*j-1})/2
\end{eqnarray}
for $n=1$ and
\begin{eqnarray}
  c(n,\alpha,j) &=& (c(n-1,\alpha,2*j) + c(n-1, \alpha,2*j-1)/2 \\
  d(n,\alpha,j) &=& (c(n-1,\alpha,2*j) -  c(n-1, \alpha,2*j-1))/2
\end{eqnarray}
for $n > 1$. The $d(n,\alpha,j)$ differ from the spatial Fourier components and the Rouse components in that they are localized; they refer to the behavior of specific parts of the polymer coil.  In contrast, the $a_{k,\alpha}(t)$ and the $C_{\alpha, n}(t)$ are both global variables, each depending on the relative positions of all the beads in the chain.

For the $a_{\alpha, k}(t)$, $C_{\alpha, n}(t)$, and $d(n,\alpha,j)(t)$ the temporal self correlation functions were evaluated.  For the $a_{\alpha, k}(t)$ and $C_{\alpha, n}(t)$, we also calculated the temporal cross-correlation functions, e.g., $\langle C_{\alpha, n}(t)C_{\beta, m}(t) \rangle$ for $\alpha \neq \beta$ and/or $m \neq n$. There are $3(N-1)$ Rouse internal modes and therefore $9(N-1)^{2}$ Rouse-Rouse self and cross-correlation functions. In Rouse's original model, if either $\alpha \neq \beta$ or $m \neq n$ or both, the temporal crosscorrelation function vanishes.

How does one show that an object is performing whole-body rotation? For a fluid velocity in the $x$ direction, with a non-zero velocity shear gradient $d v_{x}/dy$, the induced angular velocity should on the average be parallel to the $z$-axis. A simple test is advanced. If the beads are each taken to  be performing circular motion, the instantaneous angular rotation can be written
\begin{equation}
    \label{eq:angularmomentum1}
     \sum_{i=1}^{N} \bm{R}_{i} \times  \bm{v}_{i} = \sum_{i=1}^{N}  \bm{R}_{i} \times (\bm{\omega} \times \bm{R}_{i})
\end{equation}
The $z$ component of $\bm{L}$ is $\bm{\hat{k}}\cdot \bm{L}$. Applying standard identities one obtains for the rotational velocity $\bm{\omega} = \omega_{z} \bm{\hat{k}}$ around the $z$-axis
\begin{equation}
     \label{eq:angularmomentum2}
  \omega_{z} \sum_{i=1}^{N} (\langle (x_{i})^{2} \rangle + \langle (y_{i})^{2}\rangle) = \left\langle\sum_{i=1}^{N} x_{i} \frac{d y_{i}}{dt} -\sum_{i=1}^{N} y_{i} \frac{d x_{i}}{dt}\right\rangle.
\end{equation}
Corresponding forms describe rotation around the $x$ and $y$ axes. The velocities are related to the bead displacements during a single time step $\Delta t$, namely bead $i$'s displacements are $\Delta x_{i} = \Delta t \ d x_{i}/dt$, $\Delta y_{i} = \Delta t \ d y_{i}/dt$, and $\Delta z_{i} = \Delta t \ d z_{i}/dt$, so in evaluating the right-hand-side of eq.\ \ref{eq:angularmomentum2} we replace the velocities with the single-step displacements.

For whole-body rotation the two terms on the right-hand-side of eq.\ \ref{eq:angularmomentum2} are equal by symmetry. In the absence of rotation, the first sum on the right hand side of the equation will average to zero. A polymer chain is not a solid object that performs rigid-body motion, so $\bm{\omega}$ should not be overinterpreted. Sablic, et al.\cite{sablic2017a} discuss rotation in terms of Eckart frames, and note alternative definitions of rotation rates and their physical implications.

Simulations were made for 8 and 16 bead chains at shear rates $G \in (0,0.15)$;  we treat here outcomes from 16 bead chains.  In the simulations, we chose $k=1$, $f=1$, nominal temperature $k_{B}T =1$, basic time step $\Delta t = 0.001$, unit diffusion step $\Delta r = (2 k_{B}T \Delta T/f)^{1/2}$, with a unit force $k r_{i,i+1}$ giving a displacement $\Delta t/f$. The characteristic functions were computed every ten time steps. A simulation with $\Delta t = 0.0003$ gave very nearly the same results as a simulation using the longer time step.

Calculations were performed on an 8-core 3.4 Ghz CPU and an Nvidia Tesla K-40 GPU using locally written software run under Simply Fortran 2 and PGI Fortran. In production runs, polymer positions were advanced through $1 \cdot 10^{8}$ time steps. Prior to each production run, a $ 1 \cdot 10^{7}$ or longer timestep thermalization run was performed.

\section{Results}

\begin{figure}
  \centering
  \includegraphics{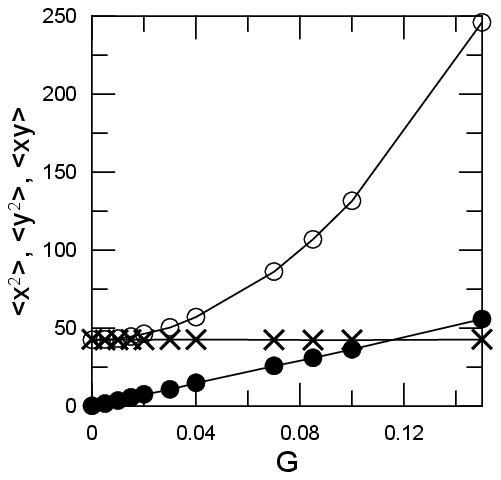}
  \caption{Effect of the shear rate $G$ on the polymer coil shape, from the equal-time correlation functions $\langle (x_{i})^{2} \rangle$ (open circles),  $\langle (y_{i})^{2} \rangle$ (crosses), and $\langle x_{i} y_{i} \rangle$ (filled circles).}
  \label{figure:nonspherical}
\end{figure}

We first consider the effect of shear on the polymer coil's shape.   As shown by Figure \ref{figure:nonspherical}, our results include both a small-shear region, in which the polymer coil is not distorted significantly, and a large-shear region, in which the polymer coil on the average is distorted by the shear.  Figure \ref{figure:nonspherical} plots the second moments $\langle x_{i}^{2} \rangle$, $\langle y_{i}^{2} \rangle$, and  $\langle x_{i} y_{i} \rangle$ against shear rate.  At zero shear, $\langle x_{i}^{2} \rangle = \langle y_{i}^{2} \rangle$.  With increasing shear, the polymer is stretched in the $x$ direction, but not in the $y$ or (not shown) $z$ directions, so that  $\langle x_{i}^{2} \rangle > \langle y_{i}^{2}\rangle$. The shear field creates a non-zero $\langle x_{i} y_{i} \rangle$ correlation that increases nearly linearly with shear rate $G$.  However, a shear $d v_{x}/dy$ has no effect on bead displacement in the $z$ direction, so  $\langle y_{i} z_{i} \rangle$ and $\langle x_{i} z_{i} \rangle$ remain equal to zero regardless of the shear rate.

We now examine the most fundamental question.  Do simulated chains rotate when placed in a shear field? Eq.\ \ref{eq:angularmomentum2} supplies the test. When the shear rate is greater than zero, the right-hand-side of eq.\ \ref{eq:angularmomentum2} is non-zero.  The polymer chair therefore rotates around the $z$-axis. We also evaluated the analogs of eq.\ \ref{eq:angularmomentum2} for rotation around the $x$ and $y$ axes.  Our shear field creates no rotation around the $x$ or $y$ axes, to within the accuracy of the simulation. For $G > 0$, rotation in the $x-y$ plane should be clockwise, i.e., $\omega < 0$, as is found.

\begin{figure}
  \centering
  \includegraphics{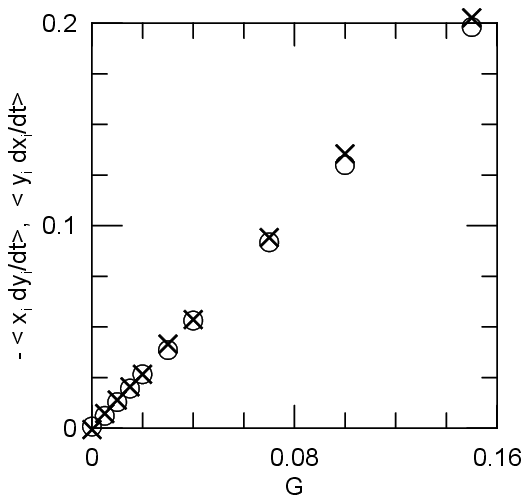}
  \caption{The $\sum_{i=1}^{N} x_{i} \ dy_{i}/dt$ ($\circ$) and $\sum_{i=1}^{N} y_{i} \ dx_{i}/dt$ ($\times$) contributions to $L$ as determined at various shear rates $G$.  As shown in eq.\ \ref{eq:rotationalhalves}, for rotational motion the two terms should be equal, as seen here to good approximation.}
  \label{figure:rotationcomponents}
\end{figure}

Is the motion actually rotational? For circular motion, the two terms on the right-hand-side of eq.\ \ref{eq:angularmomentum2} should average to the same value. Figure \ref{figure:rotationcomponents} shows that they do.  We find
\begin{equation}
       \left\langle \sum_{i=1}^{N} x_{i} \frac{d y_{i}}{dt}\right\rangle = - \left\langle \sum_{i=1}^{N} y_{i} \frac{d x_{i}}{dt} \right\rangle.
       \label{eq:rotationalhalves}
\end{equation}
Rouse predicts that the left hand side of eq.\ \ref{eq:rotationalhalves} vanishes; it does not vanish.

\begin{figure}
  \centering
  \includegraphics{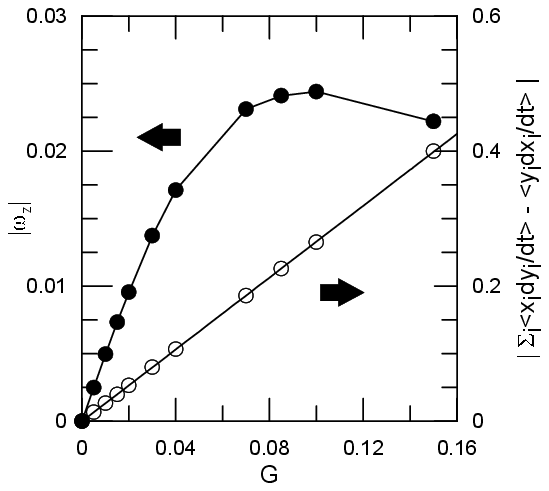}
  \caption{Nominal angular motion $|L|$ (eq.\ \ref{eq:Lvalue}, open circles) and angular rotation rate $|\omega_{z}|$ (eq.\ \ref{eq:angularmomentum2}, filled circles) for the 16-bead chain, as induced by the applied shear.}
  \label{figure:rotationrate}
\end{figure}

The nominal angular motion $L$ is
\begin{equation}
    \label{eq:Lvalue}
    L = \left\langle \sum_{i=1}^{N} x_{i} \frac{d y_{i}}{dt}\right\rangle - \left\langle \sum_{i=1}^{N} y_{i} \frac{d x_{i}}{dt} \right\rangle.
\end{equation}
Figure \ref{figure:rotationrate} gives $L$ and the rotation rate $\omega_{z}$ (eq.\ \ref{eq:angularmomentum2}), as functions of the applied shear $G$. $L$ is precisely linear in $G$ up to the largest $G$ that we examined, as predicted by Kirkwood and Riseman.  Because the polymer coil distorts when a shear is applied, $L$ and $\omega_{z}$ are not simply linearly proportional to each other.

\begin{figure}[!tbh]
  \centering
  \includegraphics{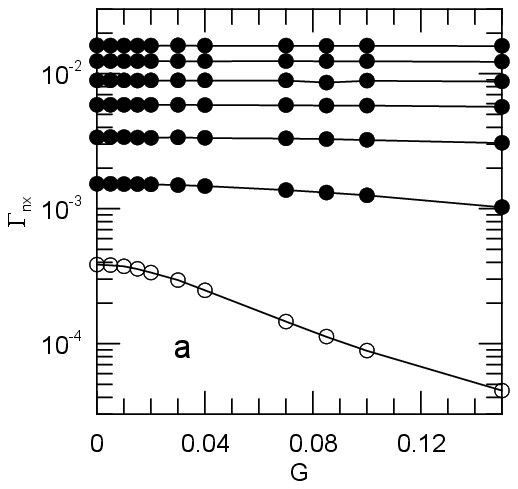}
  \vspace*{1ex}
  \includegraphics{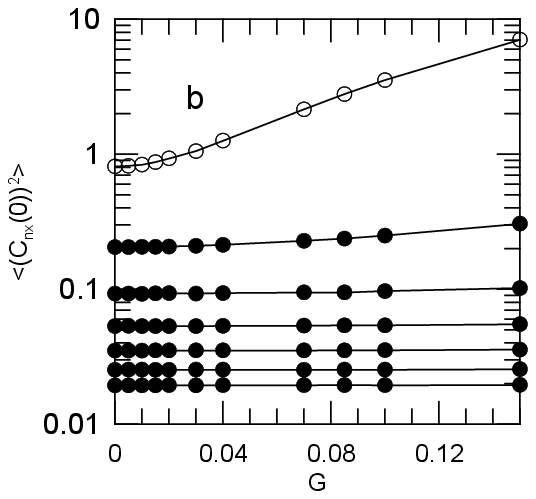}
  \caption{Effect of shear rate $G$ on the Rouse-Rouse time correlation functions  $\langle C_{nx}(0) C_{nx}(t)\rangle$, showing the dependences of (a) $\Gamma_{nx}$ and (b)$\langle (C_{nx}(0))^{2} \rangle$ on $G$.  The open circles denote the $n=1$ mode, the $n =2$ to $n=7$ modes moving seriatim away from the $n=1$ mode's behavior.  $\Gamma_{1x}$ depends on $G$ down to the smallest non-zero $G$ that we studied.}
  \label{figure:rousegamma}
\end{figure}

We now go beyond Kirkwood and Riseman, and beyond Rouse. Kirkwood and Riseman partitioned chain motions in shear into uniform translation, whole-body rotation, and residual contributions of internal modes. They ignored internal modes. In the Rouse model, mode relaxations are not perturbed by an applied shear. Here we ask whether an applied shear actually affects the internal modes, as represented by the Rouse amplitudes $C_{n\alpha}(t)$ and their relaxation rates $\Gamma_{n\alpha}$. Rouse's solutions indicate
\begin{equation}
    \label{eq:rousesolution}
     \langle C_{n\alpha}(0) C_{m\beta}(t)\rangle = \delta_{mn} \delta_{\alpha \beta} \langle(C_{n\alpha}(0))^{2}\rangle  \exp(- \Gamma_{n\alpha} t)
\end{equation}
According to Rouse's analysis: The fluctuating amplitudes $C_{n\alpha}(t)$ are uncorrelated. Modes with different $n$ fluctuate independently of each other.  Modes with the same $n$, but corresponding to different directions (different $\alpha$), also fluctuate independently. The temporal correlation function for each mode decays exponentially in time.

\begin{figure}[!tbh]
  \centering
  \includegraphics{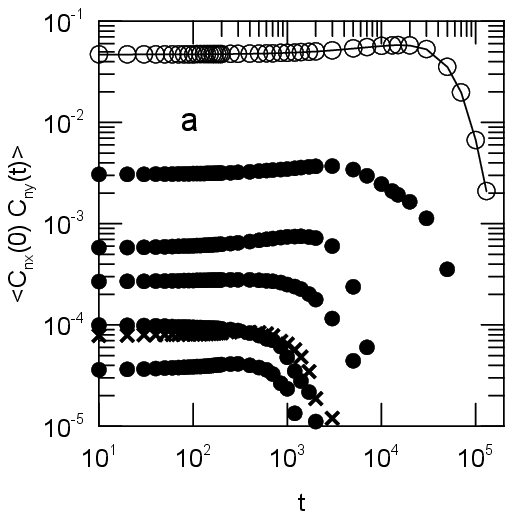}
  \vspace*{1ex}
  \includegraphics{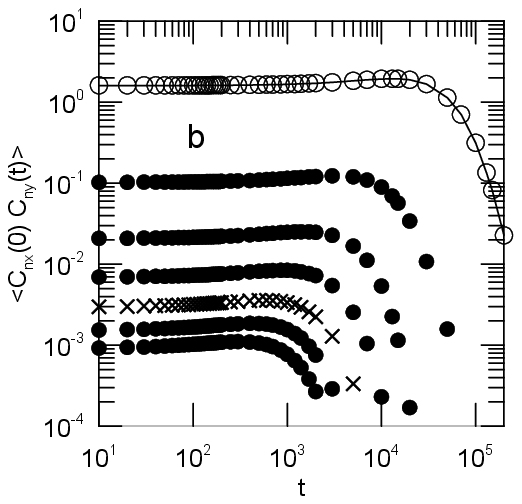}
  \caption{The Rouse-Rouse cross correlation functions $\langle C_{nx}(0) C_{ny}(t)\rangle$ for a polymer coil in shear with (a) $G=0.005$ and (b) $G=0.150$. At non-zero shear rates, the Rouse modes become cross-correlated.  Open circles mark $n=1$; crosses are (a) $n=6$ and (b) $n=5$. Note the large change in the vertical scale between these two figures. }
  \label{figure:rousecrosscorrelations}
\end{figure}

We first consider the autocorrelation functions $\langle C_{n\alpha}(0) C_{n\alpha}(t)\rangle$. We obtained the $\Gamma_{n\alpha}$ and $\langle(C_{nx}(0))^{2}\rangle$ as functions of the shear rate by fitting an early-time segment of each $\langle C_{nx}(0) C_{n\alpha}(t)\rangle$ to a single exponential. Figure \ref{figure:rousegamma}a shows the decay rates $\Gamma_{nx}$ as functions of the shear rate. Open circles mark the $n=1$ mode.  Figure \ref{figure:rousegamma}b shows the corresponding mean-square average amplitudes $\langle(C_{nx}(0))^{2}\rangle$.

\begin{figure}
  \centering
  \includegraphics{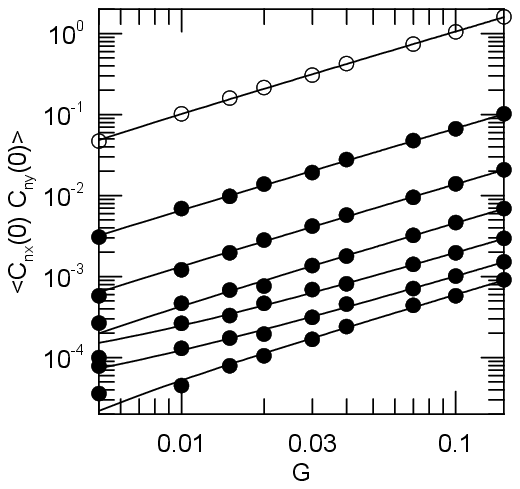}
  \caption{The initial amplitudes  $\langle C_{nx}(0) C_{ny}(0)\rangle$  of the Rouse-Rouse cross correlation functions as functions of the shear $G$.  Open circles mark the $n=1$ modes. $\langle C_{nx}(0) C_{ny}(0)\rangle$ decreases monotonically with increasing $n$. Straight lines are linear fits to $\langle C_{nx}(0) C_{ny}(0)\rangle = a G +b$, $a$ and $b$ being fitting constants.}
  \label{figure:crossamplitudes}
\end{figure}

For some modes, the decay rates and initial amplitudes are significantly shear-sensitive.  For $n=3$ and 2, and much more markedly for $n=1$, the mode relaxation rates $\Gamma_{nx}$ decrease with increasing shear rate, while the corresponding mode amplitudes $\langle(C_{nx}(0))^{2}\rangle$ increase with increasing shear rate. For $n > 3 $, $\Gamma_{nx}$ and $\langle(C_{nx}(0))^{2}\rangle$  are very nearly independent of shear rate. The relaxation rates and amplitudes for the $y$ and $z$ components of the Rouse modes are independent of the shear rate. We did not explore the dependence of this result on chain length. These non-trivial dependences of the mode amplitudes and relaxation rates on shear rate are contrary to Rouse's picture, in which the $\Gamma_{nx}$ and $\langle(C_{nx}(0))^{2}\rangle$  are not affected by solvent shear.

When shear is applied, some Rouse modes become cross-correlated. Figure \ref{figure:rousecrosscorrelations} shows the $xy$ cross-correlations $\langle C_{nx}(0) C_{ny}(t)\rangle$. These cross-correlation functions vanish in the Rouse model. They are not zero in our simulations. The corresponding $yz$ and $zx$ crosscorrelation functions (not shown) do vanish no matter whether or not shear is applied, as do all crosscorrelation functions $\langle C_{n\alpha}(0) C_{m\beta}(t)\rangle$  with $n \neq m$. The cross-correlation functions are not exponentials; they first increase and then fall off rapidly.

\begin{figure}
  \centering
  \includegraphics{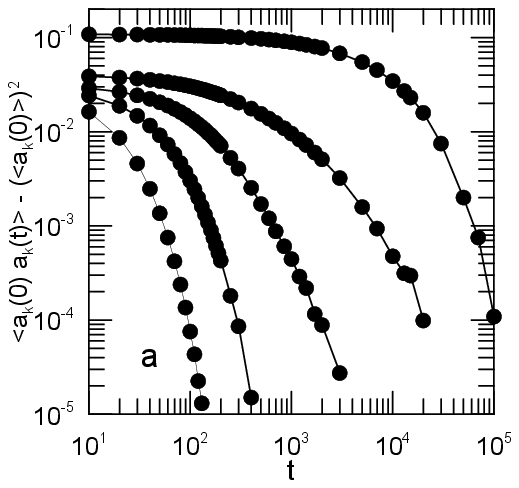}
  \vspace*{1ex}
  \includegraphics{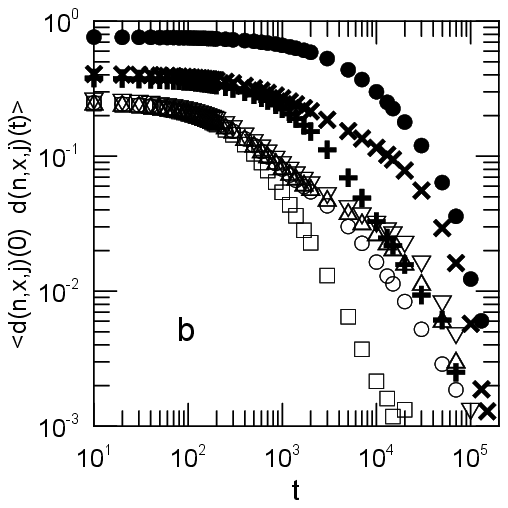}
   \caption{Temporal autocorrelation functions for (a) several spatial fourier components of the bead positions, and (b) a Haar-like wavelet decomposition of the particle positions. These representative results refer to a 16-bead chain and $G=0.03$.  The wavelet components are  $d(1,x,1)$ $(\square)$,  $d(1,x,2)$ $(\circ)$,  $d(1,x,3)$ $(\vartriangle)$,  $d(1,x,4)$ $(\triangledown)$,  $d(2,x,1)$ $(+)$,  $d(2,x,2)$ $(\times)$, and  $d(3,x,1)$ $(\bullet)$.}
  \label{figure:alternatives}
\end{figure}

The time dependences of the $\langle C_{nx}(0) C_{ny}(t)\rangle$ are qualitatively only little affected by the shear rate, but the initial amplitudes $\langle C_{nx}(0) C_{ny}(0)\rangle$ depend strongly on $G$. Figure \ref{figure:crossamplitudes} shows this dependence.  To reasonable approximation the initial amplitudes of the cross-correlation functions are linear in the shear rate $G$.

The x-y correlations are clearly driven by rotation.  Rotational motion around the $z$-axis will pump amplitude directly from $C_{nx}$ into the corresponding $C_{ny}$, as may be seen by considering the rotation of a perfectly rigid body. Whatever the amplitude $C_{nx}$ was at time 0,  at the moment the body has rotated through 90 degrees the component $C_{ny}$ is exactly equal in magnitude to the initial component $C_{nx}$. If the  $C_{nx}$ and $C_{ny}$ were initially uncorrelated, rotation will cause the cross-correlation functions $\langle C_{nx}(0) C_{ny}(t)\rangle$ to increase with increasing time. Indeed, a close examination of the cross-correlation functions in Figure \ref{figure:rousecrosscorrelations} suggests the presence of such an increase at longer times.

Figure \ref{figure:alternatives} shows representative measurements of two sets of collective coordinates that could be used as alternatives to Rouse coordinates. It is not claimed that either of these sets is necessarily the best possible choice for a set of collective coordinates, but only that there are alternatives to Rouse's coordinates that may be worth examining. The polymer coil had 16 beads; the shear rate was $0.03$. The spatial Fourier components do not decay as simple exponentials, in that they decay too slowly at longer times, but there is no sign in them of multimodal behavior.  The wavelet decompositions provide measurements of true localized motions, showing that even in this extremely simple model for polymer dynamics there is room for local differentiation of structural relaxation. We show here only the components corresponding to one-half of the full chain; the corresponding components for the other half of the chain show exactly the same set of behaviors.  In considering the series $d(1,x,j)$ (open symbols) for $j \in (1,4)$, $d(1,x,1)$, which relaxes the most rapidly, corresponds to the motions of the outer pair of beads.  The $d(1,x,j)$ for $j >2$ have clearly bimodal relaxations, speaking to more complex chain dynamics nearer to the center of the polymer. The $d(2,x,j)$ emphasize the differences between  inner and outer beads of the polymer. $d(2,x,1)$,which corresponds to the outer four beads of the polymer, has a non-exponential but unimodal relaxation; $d(2,x,2)$, describing the four beads nearest the chain center, shows a visibly bimodal relaxation.

\section{Discussion}

This paper describes a simulational study of the motions of a polymer in a shear field. Comparison was made with the Kirkwood-Riseman and Rouse treatments of the dynamics of an isolated polymer chain. We show that the Kirkwood-Riseman model of polymer dynamics, in which a polymer coil translates and rotates when subject to the influence of a shear, is qualitatively correct. The Rouse model, in which polymer coils do not rotate during viscometric studies, is incorrect as applied to the viscosity increment of a polymer in solution.  We note several alternatives to Rouse coordinates that could in principle serve as descriptions of polymer internal motions.

It is certainly legitimate to ask how the issues raised here were not already noticed. It is not suggested here that there were past errors.  Several contributory factors are readily identified. First, while there are multiple excellent presentations of the Rouse-Zimm model, e.g., refs.\ 10 and 11, equivalent presentations of the Kirkwood-Riseman model more recent than their original paper are far less common, so there is little familiarity with the Kirkwood-Riseman model. Second, in the absence of shear, the two models converge; computer simulations of polymer coils in unsheared liquids cannot readily detect the disagreement between the models.  Third, in order to identify our issues, one would have needed to analyse a chain trajectory with the correct diagnostic, e.g., eq.\ \ref{eq:rotationalhalves}, but in the context of the Rouse-Zimm model there is no rational reason to develop such a diagnostic. As a result, in past studies many fine questions have been asked about the nature of polymer dynamics, but not the questions answered here.

Larson and co-workers\cite{jain2008a,dalal2013a} provide considerable evidence that potential energies more precise than Rouse's potential can cause a chain's dynamics to deviate from simple Rouse behavior. Jain and Larson\cite{jain2008a} made Brownian dynamics simulations of a string of polymer beads to which stiff springs, bond-angle, and bond-torsion-angle forces were added seriatim.  They calculated the time autocorrelation functions for the polymer end-to-end vector and the connector unit-vector autocorrelation functions, the latter being averaged over all springs in the chain. Dalal and Larson\cite{dalal2013a} extended these results, showing that adding side groups, chain excluded-volume effects, and explicit treatment of solvent molecules jointly lead to the experimentally-observed single-exponential relaxation for short chains. They also note what they viewed as an interesting coincidence, namely that the relaxation times for the orientation of the chain end-to-end vector and the single-spring orientation vectors are very nearly the same.

The difficulty with the Rouse model is apparent on comparing eqs.\ \ref{eq:rousebeads1}-\ref{eq:rousebeads2N} with eqs.\ \ref{eq:rousebeads4}-\ref{eq:rousebeads4b}.  The second set of equations include a solvent shear force $G y_{i} \bm{\hat{i}}$ on each bead.  That force is absent from the first set of equations, the equations solved by Rouse.  Rouse calculated how a bead-spring polymer coil would evolve in time in a quiescent fluid. In a quiescent fluid, the polymer coil by symmetry has no tendency to rotate.  When a fluid shear field  $G y_{i} \bm{\hat{i}}$ is included in the calculation, the forces on the beads are changed. The motions of the beads therefore also change, and are no longer the motions described by Rouse. The Rouse model thus does not describe polymer dynamics during a rheological experiment.However, in his original paper, Rouse uses his quiescent-fluid solutions to calculate dissipation and hence viscosity increment for a polymer in a shear flow, though his solutions are not applicable under these conditions.

The above has focused on a polymer coil in an imposed macroscopic shear field, as encountered in viscoelastic measurements.  However, the fluctuation-dissipation theorem gives us two other circumstances in which polymer coils find themselves in shear fields:

First, consider a polymer coil performing Brownian motion.  The fluctuation-dissipation theorem indicates that if the chain diffuses through some distance in a given time, the chain motions will have correlated fluid motions, exactly as if the chain's motion were being created by an imposed external force. That fluid motion, the wake created in the solvent by the polymer, acts on other polymer coils, causing them to move in turn.  Because the fluid flow is not the same everywhere, those other polymer coils are subject to a fluid shear field which causes them to translate, rotate, and create fresh flow fields in the surrounding solvent. This image of flow fields being scattered and re-scattered by diffusing macromolecules forms the core of modern theoretical treatments of the diffusion of interacting spherical colloidal particles\cite{phillies2016b}, these theoretical treatments giving reasonably accurate quantitative predictions for colloidal behavior.  It should therefore not be surprising that the same general approach is valid for interacting polymer coils in solution. That is, via the fluctuation-dissipation theorem we can extend the Kirkwood-Riseman model from treating a single isolated polymer molecule to treat the hydrodynamic interactions between polymer molecules.  Indeed, there is a substantial development of polymer dynamics in non-dilute solutions based on computing the hydrodynamic interactions between polymer coils\cite{phillies2016a,phillies1988a,phillies1993a,phillies1997a,phillies1998a,phillies2002c,phillies2002b,phillies2002a,phillies2004a}.

Second, consider a polymer coil in a quiescent fluid.  On the average, there is no tendency for the molecule to rotate in any direction. However, the fluctuating thermal forces on the polymer beads create evanescent fluctuating torques on the molecule as a whole, causing the polymer end-to-end vector to perform rotational diffusion, so that its later positions gradually become decorrelated from its earlier positions.  The end-to-end vector is a sum of the individual bead-to-bead vectors, so there is a component of each bead-to-bead vector that is correlated with the chain end-to-end vector. The bead-to-bead vectors can only become completely uncorrelated on the time scale on which the chain end-to-end vector relaxes.  The result of whole-chain rotational diffusion is that the spring unit-vector correlation functions will in part relax on the time scale on which the chain end-to-end vector relaxes, precisely as found by Dalal and Larson\cite{dalal2013a}.

Rouse modes and the Rouse model are used in an extremely large number of different contexts.  I have not here generated a full list of contexts the Rouse model is inappropriate, though clearly any theoretical problem in which a polymer chain is placed in a shear field must be on that list. Nor have I considered here any extensions to the Kirkwood-Riseman model.

\end{document}